\documentclass[12pt,preprint]{aastex}
\slugcomment{Accepted, ApJ, Vol 584, 2003 February 10}
\begin{document}

\title{A Supernova Remnant Collision with a Stellar Wind }

\author{Pablo F. Vel\'azquez} 

\affil{Instituto de Ciencias Nucleares, UNAM, Apdo. Postal 70-543, CP 04510, M\'exico D.F., M\'exico\\
Email: pablo@nuclecu.unam.mx}
\author{Gloria Koenigsberger\altaffilmark{1}}
\affil{Centro de Ciencias F\'{\i}sicas, UNAM, Apdo. Postal 48-3, CP 62251, Cuernavaca, Morelos, M\'exico. \\
Email: gloria@astroscu.unam.mx}
\and
\author{Alejandro C. Raga}
\affil{Instituto de Ciencias Nucleares, UNAM, Apdo. Postal 70-543, CP 04510, M\'exico D.F., M\'exico\\
Email: raga@astroscu.unam.mx}

\altaffiltext{1}{On leave from Instituto de Astronom\'{\i}a, UNAM}

\begin{abstract}
Numerical simulations of the interaction between supernova
ejecta and a stellar wind  are presented.  We follow the
temporal evolution of the shock fronts that are formed through
such an interaction and determine the velocities, temperatures
and densities. We model the X-ray emission  from the SNR-stellar
wind collision region  and we  compare it with recent results from X-ray
observations carried out with the Chandra satellite of the SMC supernova
remnant  SNR 0057-7226 which could be interacting with the wind
of the Wolf-Rayet system HD 5980. The simulations predict the presence
of shell-like regions of enhanced X-ray emission which are
consistent with the presence of X-ray emitting arcs in the Chandra 
image. Also the observed X-ray luminosity is comparable to the X-ray 
luminosities we obtain from the  simulations for a supernova with an
initial energy in the 1-5$\times10^{50}$ erg range.  

\end{abstract}

\section{Introduction}

Massive stars are born, evolve and end their lives in clusters. Massive stars in general 
possess fast and dense stellar winds that interact with their surroundings.  Initially, 
the wind sweeps up interstellar material, until the wind-blown bubble 
encounters a 
similar bubble produced by another nearby star, at which time an interaction between these 
two stellar winds takes place.  These interactions occur at highly supersonic velocities, 
leading to diffuse emission at X-ray frequencies, as has recently been demonstrated both theoretically (Raga et al. 2001a; Cant\'o, Raga \& Rodr\'\i guez
2000)  and observationally \cite{yusef02} for the {\it Arches} cluster near the Galactic 
center.    The most massive stars have very short lifetimes ($\sim$10$^6$ years) and 
they reach the supernova (SN) stage while  all of the other stars are still on the Main Sequence, or 
have just recently evolved beyond it.  Thus, it is  to be expected that when the SN occurs,  
the ejecta will encounter and collide with the stellar winds of the other massive stars 
within the cluster.  These interactions should also be a source of X-rays.
In this paper we analyze the consequences of such  collisions, and we apply
the results to  SNR 0057-7226 in the Small Magellanic Cloud (SMC) and its possible interaction 
with the stellar wind of the Luminous Blue Variable/Wolf-Rayet (LBV/WR) binary system HD 5980.

HD 5980 is the most luminous point source in the Small Magellanic Cloud and is believed to 
consist of a close  binary pair, both having Wolf-Rayet characteristics 
(Niemela 1988) and a third, possibly line-of-sight O4-6 type  
component (Koenigsberger et al. 2002).  Its eruptive mass-shedding events
as well as the wind-wind interactions within the close pair have kept it
under scrutiny for several years. HD 5980 is found at the edge of NGC 346, the 
largest cluster in the Small Magellanic Cloud. The age of the cluster is estimated 
at $\sim$10$^6$ years.

A nebulosity believed to be a supernova remnant lies on the foreground of HD 5980.
This remnant might give rise to the $+300$~km~s$^{-1}$ component observed in the
UV absorption lines towards HD 5980 (see Fitzpatrick \& Savage 1983).
This nebulosity, SNR 0057-7226, is associated with the X-ray source IKT 18
(Inoue 1983), that was confirmed to be  a non-thermal radio shell by Ye et al. (1991), 
having a diameter of 3\rlap{\arcmin}.2~(55 pc). It is the second
most luminous SNR in the SMC 
at 843 MHz.  An X-ray image recently obtained with Chandra \cite{naze02} displays 
extended emission ($130\arcsec \times 100\arcsec$; i.e., 39$\times$29~pc) 
with a luminosity of
$1.40\times10^{35}$ ergs s$^{-1}$ in the 0.3-10.0 keV band.  The fact that
the X-ray emission is centered on HD 5980 could well be a coincidence. However, 
UV and FUV observations (de Boer \& Savage 1980; Koenigsberger et al. 2001; 
Hoopes et al. 2001) indicate 
that the portions of the SNR that are approaching the HD 5980 system have a different 
ionization balance than the opposite side (i.e., the side  approaching the observer).
These observational results could be explained if the receding portions 
of the SNR are interacting with the stellar wind from  HD 5980.
The analysis of this possibility is the primary motivation for undertaking
the present study.

\section{Model for the SNR-stellar wind interaction}

\subsection{The yguaz\'u-a code}

For the numerical simulations carried out in this paper, we have used
an axisymmetric version of the {\it yguaz\'u-a} adaptive grid 
code, which is described in detail by Raga et al. (2000) and which has
been tested with laboratory explosion simulations
(Sobral et al. 2000;
Vel\'azquez et al. 2001a), in SNR models \cite{vel01b} and
jets \cite{raga01b}. The gasdynamic equations are integrated with a 
second order accurate (in space and time) implementation of the 
``flux-vector splitting" algorithm of Van Leer (1982), together with
a system of rate equations for atomic/ionic species.

\subsection{Initial conditions and assumptions}

Due to the symmetry of the problem given by the axis which is joining the 
location of the SN explosion with the position of the star, we 
perform the integration of the cylindrically symmetric gasdynamic equations.
 A 4-level binary adaptive grid with
a maximum resolution of $1.46\times10^{17}$~cm was employed, in a 
50$\times$~25 pc (axial$\times$~radial) computational domain.

For the initial conditions, several assumptions are possible: (a) the SNR and 
the stellar wind are ``turned on" simultaneously, (b) the stellar wind is 
already present when the SN explodes, or (c) the SN explodes first, and then 
the stellar wind is ``turned on". 
We assume that the SN explosion and the stellar wind are turned on at the same
time. 

Simulations can be carried out varying all of the different parameters
of the calculation (terminal wind velocity, mass loss rate, distance 
between the star and the SNR, energy of the SN explosions, ISM density). 
However, we limit the present study to models with  characteristics 
applicable to the HD 5980 system. Hence,  in all simulations we have 
fixed the stellar wind mass-loss rate at  
$\dot M=5\times 10^{-6}~M_{\odot}$~yr$^{-1}$, which is typical of
massive O-type stellar winds. The wind-shedding star is located  at
$3\times10^{19}$~cm from the origin of the coordinate system, while 
the center of the SNR is located at $8\times10^{19}$~cm. Thus, the 
separation between the pre-SN and the normal star is of $5\times10^{19}$~cm.
The calculations were initialized considering that the the radius of the
initial remnant is of $4\times10^{18}$~cm.  The density of the homogeneous
ISM into which the SNR and the stellar wind are expanding is assumed to be
of 1~cm$^{-3}$, while its temperature is of $10^{4}$~K.

Five numerical simulations were carried out considering different values for 
the initial SN explosion energy, $E_0$, and the terminal velocity of the
stellar wind, $v_{\infty}$. These numerical models are labeled
``A'' to ``E'' and  are listed in Table \ref{table1}. The different values
of $E_0$ are $10^{50}$, $5\times 10^{50}$, and $10^{51}$ ergs, while the values
of $v_{\infty}$ are 1500 and 2500 km s$^{-1}$. Columns 2
and 3 of Table \ref{table1} list the values of these parameters that 
correspond to each model.

In order to compare our numerical simulation with Chandra observations,
we have computed X-ray emission maps for the  0.3-10 keV photon energy range, 
using the  CHIANTI\footnote{The CHIANTI database and associated IDL 
procedures are 
freely available at the following addresses on the World Wide Web: 
http://wwwsolar.nrl.navy.mil/chianti.html, http://www.arcetri.astro.it/science/chianti/chianti.html, 
and http://www.damtp.cam.ac.uk/users/astro/chianti/chianti.html.} atomic 
database (Dere et al. 2001).  For the calculation of the X-ray emission
coefficient we have assumed a modified Shull ionization equilibrium 
(Shull \& van Steenberg 1982; Arnaud \& Rothenflug 1986; 
Landini \& Monsignori Fossi 1991) 
and that all of the excitation is in the low-density regime 
(in this case, the  emission coefficient is proportional to the square 
of the density). The abundances for heavy elements were taken from the fit
to the X-ray spectrum by Naz\'e et al. (2002); i.e., $Z=0.17\ Z_{\odot}$.

\section{Results}

\subsection{Evolution of the collision between a SNR and a stellar wind}

Figure \ref{fig1} shows a comparison between the density evolution for
model A (panels on the right) and model E (panels on the left). 
The top panels (for $t=8000$~yr and $t=17000$~yr) represent the density
distributions at the time when the encounter between the SNR shock wave
and the stellar wind first occurs. The spherical shell located on the left 
hand side is the wind-blown bubble surrounding the star. Note that there
is a difference of more than 2 orders of magnitude between the density in
the wind-blown bubble and the stellar wind which it encloses, due to the
fact that the stellar wind has swept up material from the interstellar medium
into which it is flowing. As expected, the more energetic SN results in a remnant 
that expands and reaches the wind of the star sooner (8000 yr, model E) than the less 
energetic SN (17000 yr, model A).

Once the SNR shock wave has reached the wind-blown bubble, 
two secondary shock waves are produced, one propagating into the remnant (S1)
and the second one in the direction of the star (S2), sweeping up the material
in the wind-blown bubble and in the stellar wind. This is illustrated in
the following panels of Figure 1, which present the density evolution at
temporal intervals of 5000 years, from top to bottom. 
The secondary shock wave that is
propagating into the SNR material (S1) is almost spherical, unlike  the shock
wave that is  moving into the stellar wind (S2), which  initially presents a wavy
structure that subsequently evolves into finger-like features.  The propagation of these
finger-like structures is clearly seen in model E,
in which the evolution occurs more rapidly than in the lower energy SN models.
At $t\sim 23000$~yr (bottom left panel), these structures have
reached  the opposite side  of the wind-blown bubble.  The eventual fate 
of the wind-blown bubble is that only a small portion lying close to the 
symmetry axis of the interaction remains while the rest will have been swept up 
by the SNR. As the SNR advances into the stellar wind, the S2 shock adopts the shape 
of a bow-shock, enveloping the star.  The bottom-left panel of Figure 1 shows this
structure very clearly (at $x\sim 3.5\times10^{19}$~cm) .

The finger-like features are  most likely produced by Rayleigh-Taylor (R-T) 
instabilities for the following reasons. The basic conditions for the onset 
of the R-T instability are the existence of two media with different densities, 
and  an effective gravity with an orientation in the direction that goes from 
the high density to the low density medium.  Both of these conditions are met 
in the interaction region. The gas behind the SNR shock wave is almost 
two orders of magnitude denser than the gas in the wind bubble; and a 
calculation of the gradient of the ram pressure across the interaction 
region for the $E_0=10^{51}$~erg simulation at $t=13000$~yr, yields 
an effective gravity at the location  of the SNR/wind interaction region 
of $\sim 6\times 10^{-4}$~cm s$^{-2}$, directed outwards from the SN.
Furthermore, using the values given above for the effective gravity and the
density ratio, and considering a typical size of $\sim 10^{19}$~cm
for the initial perturbation (consistent with the radius of the wind
bubble at the time when the wind/SNR interaction starts to take place,
see Fig. \ref{fig1}), one obtains a $\gamma=1.94\times 10^{-11}$~s$^{-1}$ growth 
rate for the R-T instability. This growth rate is equivalent to a $\tau=1600$~yr
timescale, which is consistent with the time in which the finger-like
structures develop in our simulation (which is smaller than the 5000~yr 
time-intervals between the successive frames of Fig. \ref{fig1}). 

The qualitative nature of the interaction does not depend significantly on the
velocity of the stellar wind, for values that are reasonable for O-type stars.
As illustrated in Figure \ref{fig2},  the  general features described above for the
density evolution  of the interaction also appear for a stellar wind
having $v_{\infty}=1500$~km s$^{-1}$ (model B, panels on the left  of Fig. \ref{fig2}).  
The panels on the right of this figure reproduce the case with 
$v_{\infty}=2500$~km s$^{-1}$ (model A).  From top to bottom, the panels
of Figure 2 display the morphology of the interaction at  temporal
intervals of 10000~yr, starting with initial contact between the SNR and the
stellar wind.

An example of the kinematical characteristics of the flow that would be seen by an 
observer located along the symmetry axis in the positive $x$ direction
is shown in Figure \ref{fig3}.  Assuming that the observer is at
rest with respect to the undisturbed environment of our simulations,
the radial velocities of the absorbing (or emitting) gas are
given by $v_{obs}=-v_x$ (where $v_x$ is the axial velocity of the flow).
In this example, we present the velocity profile  of the gas that lies 
along the symmetry axis at t=27000~yr, for model A 
(see table \ref{table1}). The wind-shedding star is located at 
$x=3\times10^{19}$ cm,
and the portion of its stellar wind that is receding from the observer can
be seen as the (cut-out) maximum to the left of this position.  The portion of
its stellar wind that is approaching the observer is seen as the (cut-out) minimum,
just to the right of this position. The extension in the $x$-direction of this
portion of the stellar wind is not as great as in the case of the receding wind,
since it encounters the SNR ejecta (denoted as ``a'' in Fig. \ref{fig3}).  
This encounter not only slows down the wind, but it leads to a shock wave (S2, described above)
having a velocity in the opposite direction (i.e., towards the star).    
It is interesting that even though the expansion velocity of the SNR is lower 
than the velocity of the stellar wind, the balance of the momentum is in 
favor of the SNR and the interaction region has this net velocity towards the star. 
The velocity  along the $x-$axis towards the star at position ``a'' is 
120 km s$^{-1}$, not too different from the radial velocity of high-ionization 
ultraviolet ISM components observed in absorption at 150 km s$^{-1}$, corresponding to $v_{hel}\simeq 300$~km s$^{-1}$ after sustracting the average 
systemic velocity for the SMC of $\sim$ 150~km s$^{-1}$
(de Boer \& Savage 1980; Koenigsberger et al. 2001), and the \ion{O}{6}
components observed in the far-ultraviolet absorption components (Hoopes et al. 2001).

The shock wave that is moving  into the SNR (S1) has a velocity  
of $\sim$$-$40 km s$^{-1}$ (denoted as ``b'' in Fig. \ref{fig3}) in the direction of 
the observer. In front of this shock, unperturbed SNR material is found, having
velocities of 60 km s$^{-1}$, decreasing systematically as the location 
of the progenitor of the SNR (at $x=8\times 10^{19}$ cm) is approached, where $v_x=0$. 
The velocities between this point and the unperturbed ISM, in the direction of the observer, range from 0 to $-$150 km s$^{-1}$. Note that this latter velocity, when corrected for the
average SMC systemic velocity ($\sim$ 150 km s$^{-1}$), lies  at the velocity of the
local (galactic) ISM.  Hence, the absorption features that may be produced in this
region of the SNR (marked ``c" in Fig. 3) would be superimposed on the local
Galactic ISM absorptions and thus would be very difficult to observe. 

If we assume a systemic velocity of $v_{snr}\simeq 188$~km s$^{-1}$ for
the SNR (Koenigsberger et al. 2001), the absorption components generated by 
our models at the opposite 
sides of the SNR (locations ``a'' and ``c'' in Fig.~\ref{fig3}) would be 
observed at $v_{hel}=(+120+188)$~km s$^{-1} =+308$~km s$^{-1}$ and 
$v_{hel}=(-150+188)$~km s$^{-1}=+38$~km s$^{-1}$, thus suggesting that the pair of lines
corresponding to the SNR that were observed by Koenigsberger et al. (2001)
are really the $v_{hel}=(+313,+33)$~km s$^{-1}$ pair.  

\subsection{Simulated X-ray emission}

For all of the numerical models we have obtained simulated X-ray emission maps.
For each of these maps, we have calculated total X-ray luminosities 
within the 0.3-10 keV band at different times, and these results are 
summarized in columns 5 and 6 of Table \ref{table1}.  The time for which
the calculation is made is listed in column 4.  The first time corresponds
to initial contact between the SNR and the stellar wind; the second and
third times correspond in all cases, except for model E, to 10000 yrs and
20000 yrs after initial contact.  In the case of model E, the calculation
was made at 15000 yrs.

The relative orientation of the SNR/wind collision region with respect
to the observer is an important parameter for the calculation of the 
spatial distribution of the X-ray emission, and we shall discuss this below.
For the moment, however, we shall concentrate on the case in which  the 
angle between the $x-$axis of our calculations and the line-of-sight to 
the observer is 0\arcdeg;  that is, the observer is situated on the extreme
right in Figures 1 and 2 and, therefore, views the interaction region
``face-on''.

The top panel in Figure \ref{fig4} displays the X-ray emission
at the time of initial contact between the SNR and the stellar wind for 
model A. Superposed 
on the emission arising in the SNR as a whole, two zones of higher intensity 
X-ray emission can be identified: a) the  rim of the SNR, 
where the mass density is high due to the swept-up ISM; and b) a central dot, which 
corresponds to the location of first contact between the SNR and the wind.  
The X-ray emission at this stage is dominated by the SNR. The total X-ray
luminosity is $7.9\times 10^{34}$~erg s$^{-1}$ in the 0.3-10 keV band, of
which only 1.3$\times$10$^{34}$~erg s$^{-1}$ arises in the shocked and 
unshocked stellar wind.

As time procedes, the interaction between the SNR and the stellar wind 
involves more and more material, and the emission from the unperturbed SNR 
decreases. Thus, the relative  contribution to the X-ray emission arising in 
the interaction zones becomes more significant. The second panel in Figure \ref{fig4}
illustrates the X-ray emission at $t=37000$ yr, 20000 yr after initial contact.
The ring of intense X-ray emission that is visible in this figure corresponds 
to the regions of stellar wind that are making their first contact with the 
SNR. These regions have densities that are particularly high due to the 
presence of swept-up material (both by the SNR and the stellar wind).  
There is also a small  ring near the center of the bright ring which, 
although significantly fainter,  is visible in this same figure.  This ring 
corresponds to  emission from the bow shock that by this time has been formed,
enveloping the star. The total X-ray luminosity at this time is 
$3.9\times 10^{34}$~erg s$^{-1}$.  Our simulations indicate that the
combination of unperturbed stellar wind + shocked stellar wind contribute 
1.4$\times$10$^{34}$~erg s$^{-1}$ to this luminosity; i.e., 35\% 
of the total X-ray luminosity.

Figure \ref{fig5} presents the same results discussed above in relation with 
Figure \ref{fig4}, but for model E. The outer ring and the inner ring 
correspond to the interaction of the SNR with the ISM and the bow shock
enveloping the star, respectively, as in Figure \ref{fig4}.  In addition, 
the third ring (counting outwards) is analogous to the bright ring in 
the lower panel of Figure \ref{fig4}; i.e., it corresponds to the first contact 
between previously unperturbed portions of the wind-blown bubble and the SNR.
Because the SN in this example is more energetic, the evolution of 
the SNR/wind interaction takes place on shorter timescales than in the example
shown in Figure \ref{fig4}. Thus, the lower panel in Figure \ref{fig5}
displays features that have not yet formed at the evolutionary time 
illustrated in the lower panel of Figure \ref{fig4}. In particular, the  
second ring (counting from the inside) corresponds to the
finger-like features illustrated in the right panels of Figure \ref{fig1}.  
The total X-ray luminosities are much larger for this case, reaching values of up 
to 9$\times 10^{35}$~erg s$^{-1}$ at $t=23000$ yr, but the stellar wind+interaction
regions account for only $10^{34}$ ergs s$^{-1}$ of this emission.

These simulated X-ray luminosities may  be compared with the values 
obtained from the Chandra maps of the regions around HD~5980 by Naz\'e et al. 
(2002), who derive $L_x=1.4\times10^{35}$ erg s$^{-1}$ in the 0.3-10 keV band. 
This luminosity lies within the range of values for the luminosity of the
entire region (SNR+stellar wind+interaction regions; column 5 of Table \ref{table1})
obtained from our simulations, for a SN explosion with $E_0$ in the 
1$-$5$\times$10$^{50}$ erg range. For $E_0 \ge5\times$10$^{50}$ ergs,
the predicted X-ray luminosities are too large compared with the observations.
Note that for such an energetic SN, the relative contribution to the X-ray luminosity
is completely dominated by the SNR, with a negligible contribution from the
stellar wind and from the wind/SNR interaction region. 

We now turn to the morphology of the region, and as mentioned above, the
angle between the symmetry axis ($x-$axis) and the plane of the sky as viewed
by an external observer has a significant effect on the shape of the X-ray 
emitting region. We have already discussed the morphology for the case in which
the observer is located along the $x-$axis. In Figure \ref{fig6}, we compare 
the
spatial distribution of the X-ray emission arising in the density
distribution computed for model A at 37000 yrs and for model E at 23000 yr,
 but viewed at three different angles.  At 80\arcdeg, the morphology is very 
similar to that illustrated in Figures 4 and 5, but the central rings are displaced 
to one side of the star's location.  At  45\arcdeg, the displacement is significantly
more noticeable, and in the case of model E, only a segment of the 
bright ring formed by the interaction between the stellar wind and the SNR 
is apparent. At 60\arcdeg, the morphology is intermediate between the other 
two angles for which the X-ray emission is illustrated. 

Naz\'e et al. (2002) describe the morphology of the X-ray emission as having a
size of $130\arcsec \times100\arcsec$, i.e. 37$\times$29~pc and  containing  a few bright 
or dark arcs, with no obvious limb-brightening.  One of the arcs that appears to 
be centered on the star is rather diffuse and clumpy, and has an approximate diameter 
of 60\arcsec, corresponding to a physical size of  $\sim {5\times 10^{19}}$~cm.  In
Figure 7 we reproduce the X-ray image of HD 5980, presented by Naz\'e et al. (2002; their
Figure 8), and compare it with the results of model A (top panel;
t=37000 yr) and model E (middle panel; t=23000 yr), both of which are calculated
under the assumption of an 80\arcdeg angle between the symmetry axis and the plane of
the sky.  The crosses in the simulations indicate the location of 
the star, which is clearly visible as the dark dot in the Chandra image
(bottom panel).  The size of the extended emission is closer to that given by
model A than that of model E.  If we assume that the outermost regions of
the extended emission can be associated with the SNR rim, then this region
is also more similar to model A than to model E, which has too bright a
rim.  However, the contrast between the internal ring in model A and the
background X-ray emission is much greater than the observed contrast.  
In this respect, the observations are more similar to model E.  
Furthermore, there is obviously much more structure in the observations as 
predicted by model E, rather than model A.  Hence, the extended emission associated with
HD 5980 presents a morphology that may be described as intermediate between
the morphologies predicted by models A and E, consistent with the result
that the observed X-ray luminosities lies between the ones predicted from
model A and E.  A feature that is not
predicted by the simulations is the protrusion at the top left of the
X-ray image.  Clearly, the simulations are based on idealized assumptions
concerning the symmetry and homogeneity of both the SNR and the stellar wind
and its wind-blown bubble.  In reality, each of these components are likely
to be clumpy and may depart significantly from spherical symmetry, and could
possibly give rise to the more complex structures observed in the Chandra
image.

\section{Discussion}

In this paper we describe the interaction of a SNR with the wind of a nearby massive star.  
This interaction leads to localized sources of X-ray emission embedded in the diffuse emission of the SNR. The emission arising in the SNR/stellar wind interaction is a
substantial fraction of the total X-ray luminosity several thousand years after the 
SNR-stellar wind interaction has initiated.  The X-ray luminosity depends primarily
on the energy of the supernova event.  The simulations are consistent with the
general features recently observed in the extended X-ray emission surrounding
the WR/LBV binary system HD 5980, supporting the possibility that this extended
X-ray emission includes contributions from the interaction between the SNR and
HD 5980's wind.  Assuming this scenario, the initial energy of the SN was
1-5$\times^{50}$ ergs, and the SN event occurred between 23000 and 37000 years
ago.  In support of this scenario, the radial velocities of ISM features that have been
observed in the UV and FUV wavelength regions at $v_{hel}\simeq 300$ km s$^{-1}$ (or $v\sim$150 km s$^{-1}$ with respect
to the stationary SMC ISM) are consistent with the velocity of the shock front
that is expanding into the stellar wind of HD 5980  which is predicted by the 
simulations. However, before concluding that the 300~km~s$^{-1}$ heliocentric
component
observed in the spectrum of HD 5980 indeed comes from the wind/SNR
interaction region,  detailed predictions of the absorption line profiles
using gasdynamic models are required.

The simulations predict the presence of X-ray bright rings.  The observations
indicate the presence of arcs, rather than rings.  The rings appear as a
consequence of the assumed symmetry in the stellar wind and in the SNR.
Clearly, a more detailed model for HD 5980 should contemplate the
possibility of a more irregular structure.  In particular, HD 5980 consists
of two stars with strongly interacting stellar winds, which can lead to
the formation of large-scale structures.  In addition,  HD 5980 has undergone 
a recent LBV-type eruption associated with the star that is currently 
undergoing a transition from O-type to WR star. In the past,  similar 
enhanced mass-loss episodes are likely to have occurred.
Hence, the stellar wind  which the SNR is encountering
is likely to be structured, with high and low density regions.  This will lead
to additional fluctuations in the density stratification of the interaction
region, and will definitely produce a more irregular X-ray emission map 
than the one predicted from our simple model.   

The interaction with the SNR ends up destroying the wind-blown bubble that 
surrounds the star, leaving only a bow shock separating the SNR and the 
stellar wind, and a small segment of the material that was swept-up by the 
stellar wind on the opposite side of the star. The entire phenomenon for the 
cases we have modeled is very short-lived  ($\sim 10^4$~yr), which, combined 
with the low frequency of SN events, makes it a very rare phenomenon, not  
likely to be frequently observed.  However, a long-lasting effect is the fact 
that once the interaction of the SNR with the stellar 
wind has transpired, any material that had been swept up by
the stellar wind is removed, and the subsequent propagation of the stellar 
wind into the ISM proceeds unhindered, as it expands into the SNR cavity.  

The short-lived phenomenon we describe must occur in all young massive star clusters.
We can speculate that if the SN explodes near the center of the cluster, the ejecta 
will interact with a large number of O-type stars, not just one, as in our simulations.
Assuming that each interaction region leads to a luminosity of $\sim$10$^{35}$ ergs s$^{-1}$,
and that there are 100 nearby O-type stars, a luminosity of $\sim$10$^{37}$ 
ergs s$^{-1}$ is predicted from the whole cluster.

\acknowledgements

We thank Ian Stevens and Mike Corcoran for reading and commenting on the 
draft of this paper, and Yael Naz\'e for authorizing the reproduction of
part of Figure 8 from  Naz\'e et al.~(2002).  We also thank  an anonymous referee 
for numerous comments and suggestions that helped to significantly improve 
the paper.
PV and AR are supported by CONACYT grants 34566-E and 36572-E. GK acknowledges
financial support from CONACYT grant 36569-E. We thank Israel D\'\i az for
computer support.

\clearpage

\begin{deluxetable}{ccccccc}
\tablecaption{Models and simulated X-ray luminosities\label{table1}}
\tablewidth{0pt}
\tablehead{
\colhead{Model} & \colhead{$E_0$} & \colhead{$v_{\infty}$}& 
\colhead{$\tau$~\tablenotemark{a}}   & \colhead{$L^{snr+w}_x$~\tablenotemark{b}}   &
\colhead{$L^{w}_x$~\tablenotemark{c}} & \colhead{\% of $L^{w}_x$/$L^{snr+w}_x$}  \\
\colhead{}  & \colhead{10$^{51}$~erg}  & \colhead{km s$^{-1}$}  &
\colhead{$10^3$~yr}   &  \colhead{$10^{34}$~erg s$^{-1}$} & 
\colhead{$10^{34}$~erg s$^{-1}$} & \colhead{}
}
\startdata
 A & 0.1 & 2500 &  0. & 7.9  & 1.3 & 16. \\
 A &  &  & 10. & 5.6  & 1.7 & 30. \\
 A &  &  & 20. & 3.9  & 1.4 & 35. \\
 B & 0.1 & 1500 &  0. & 5.7  & 1.3 & 23. \\
 B &  &  & 10. & 3.8  & 1.1 & 28. \\
 B &  &  & 20. & 2.6  & 0.6 & 22. \\
 C & 0.5 & 2500 &  0. & 44.3 & 0.35 & 8.6$\times 10^{-2}$\\
 C &  &  & 10. & 43.2 & 0.73 & 1.7\\
 C &  &  & 20. & 37.8 & 2.6 & 7.0 \\
 D & 0.5 & 1500 &  0. & 44.3 & 0.04 & 8.6$\times 10^{-2}$\\
 D &  &  & 10. & 42.2 & 0.08 & 0.18\\
 D &  &  & 20. & 34.4 & 0.17 & 0.5\\
 E & 1.0 & 2500 &  0. & 78.0 & $\sim 0$ & $\sim 0$\\
 E &  &  & 15. & 93.0 & 1.0 & 0.8 \\
\enddata
\tablenotetext{a}{$\tau=t-t_{0}$, where $t$ is the time and $t_0$ is
the time when the interaction between the SNR and the stellar wind bubble
first ocurrs.}
\tablenotetext{b}{X-ray luminosity from SNR plus the stellar wind and its interaction.}
\tablenotetext{c}{X-ray luminosity from the interaction region and 
stellar wind only.}
\end{deluxetable}

\clearpage


\section{Figure captions}
\figcaption{Temporal evolution of the density distribution for the interaction
between a SNR shock wave (large bubble on the right of each panel) and a 
stellar wind (small bubble on the left), for  model A 
($E_0=10^{50}$ ergs; right panels) and model E 
($E_0=10^{51}$ ergs; left panels).  The reflected shock waves produced
after the interaction are labeled  S1 (which is moving into SNR material)  
and S2 (moving into stellar wind bubble gas).  The greyscale
is logarithmic in the  $10^{-26}$--$10^{-22}$ g cm$^{-3}$ range (given 
by the vertical bar on the right of bottom panel).
The distance scale in both axes is given in cm.
\label{fig1}}


\figcaption{The same as Figure \ref{fig1}, but for  model A 
($v_\infty$=2500 km s$^{-1}$; panels on the right) over longer time
scales and model B ($v_\infty$=1500 km s$^{-1}$; panels on the left). 
The small shell (on the left of each frame) is the stellar wind bubble.
The evolution in both cases is quite similar.
\label{fig2}}


\figcaption{Observed gas velocity $v_{obs}$ along the symmetry axis 
for model A, at a $t=27000$~yr time. The observer is situated 
on the symmetry axis, to the right of the SNR (see Figs. \ref{fig1} 
and \ref{fig2}). The letter ``a'' indicates the position and velocity of 
gas moving towards  the star; ``b''  the location and velocity of
the shock wave moving into the SNR; and ``c'' the position and velocity 
of the SNR shock wave which is not interacting with the stellar 
wind. The star is located at $x=3\times 10^{19}$~cm and the 
center of the SNR is at $x=8\times 10^{19}$~cm. The horizontal 
axis is the same as in Figures \ref{fig1} and \ref{fig2}.
\label{fig3}}


\figcaption{Maps simulating the X-ray emission at times $t=17000$
and 37000 yr, for model A, obtained by integrating the 
emission coefficient  in the energy range 0.3-10 keV along lines of sight. 
The greyscales are linear and are given by the bars to the
right of both maps. The X-ray emission is given in units of erg s$^{-1}$
cm$^{-2}$ sterad$^{-1}$.  The angle between the symmetry axis ($x-$axis on 
Figs.~\ref{fig1} and \ref{fig2}) and the plane of sky is of 90\arcdeg~
(i.e., the observer is located to the right of the SNR, on the line which 
is joining the star and the SNR center, see Figs..~\ref{fig1} and \ref{fig2}).
 A comparison between the two maps shows that 
the X-ray emission from the interaction region becomes important 
at later times after the interaction has initiated,  as illustrated by
the bright ring in the $t=37000$~yr frame.
\label{fig4}}


\figcaption{Same of Figure \ref{fig4}, but for  model E. 
Several X-ray emission rings are observed  at $t=23000$~yr, resulting 
from the various interactions described in the text (edge of interaction 
region, shock between the dense finger-like structure and the receding side 
of the stellar wind bubble, and the bow shock).
\label{fig5}}


\figcaption{Dependence on the angle $\phi$ (between the plane of the sky
and the $x-$axis) of the X-ray distribution for  model A at $t=37000$ yr
(panels on the left), and model E at $t=23000$~yr (panels on the right).
\label{fig6}}

\figcaption{Comparison of the predicted morphology in model A (top panel;
$t=37000$~yr) and model E (middle panel; $t=23000$~yr), both viewed at a
$\phi$=80\arcdeg~angle, with the Chandra image (from Naz\'e et al. 2002,
with permission from the authors; bottom panel) of the extended X-ray 
emission associated with HD 5980.
\label{fig7}}


\begin{thebibliography}{}


\bibitem[Arnaud \& Rothenflug 1986]{ar86} Arnaud, M., Rothenflug, R., 1986, A\&AS, 60, 425.            
\bibitem[Cant\'o, Raga \& Rodr\'\i guez 2000]{canto00} Cant\'o, J., Raga, A.C. \& Rodr\'\i guez, L.F. 2000, ApJ, 536, 896.
\bibitem[de Boer \& Savage(1980)]{boer80} de Boer, K.S., \& Savage, B.D. 1980, ApJ, 238, 86.
\bibitem[Dere et al. 2001]{dere01} Dere, K.P., Landi, E., Young, P.R. \& Zanna, G. 2001, ApJS, 134, 331.
\bibitem[Fitzpatrick \& Savage (1983)]{fitz83}Fitzpatrick, E. L. \& Savage, B.
D. 1983, \apj , 267, 93.
\bibitem[Inoue, Koyama \& Tanaka 1983]{inoue83} Inoue, H., Koyama, K  \& Tanaka, Y. 1983, In: Supernova Remnants
     and their X ray Emission, IAU Symp. No. 101, p. 535, eds Danzinger, J.I. 
     and Gorenstein, P., (Reidel: Dordrecht).
\bibitem[Hoopes et al. 2001]{hoopes01} Hoopes, C.G., Sembach, KR., Howk, J.C. \& Blair, W.P.. 2001, ApJ, 558, L35.
\bibitem[Koenigsberger et al. 2001]{koe01} Koenigsberger,G.,Georgiev, L., Peimbert,M.,Barb\'a, R., Niemela, V.S., Morrell, N., Walborn,N.R., Tzvetanov, Z., \& Schulte-Ladbeck, R. 2001, AJ, 121, 267. 
\bibitem[Koenigsberger, Kurucz \& Georgiev 2002]{koe02}Koenigsberger, G., Kurucz, R., Georgiev, L. 2002, ApJ, in press. 
\bibitem[Landini \& Monsignori Fossi 1991]{landini91} Landini, M., Monsignori Fossi, B.C., 1991, A\&AS, 91, 183.
\bibitem[Naz\'e et al. 2002]{naze02} Naz\'e, Y., Hartwell, J.M., Stevens, I.R., Corcoran, M.F., Chu, Y.-H.,
           Koenigsberger, G., Moffat, A.F.J., and Niemela, V.S. 2002, ApJ, in press.
\bibitem[Niemela 1988]{niemela88} Niemela, V.N. 1988,  in Progress and Opportunities in Southern Hemisphere Optical Astronomy, ed. V.M. Blanco \& M.M. Pillips (Provo, UT: Brigham Youn Univ. Press), 381.
\bibitem[Raga et al. 2002]{raga02} Raga, A.C., de Gouveia dal Pino, E., Noriega-Crespo, A., 
 Mininni, P.D. \& Vel\'azquez, P.F. 2002 , A\&A, 392, 267.
\bibitem[Raga et al. 2001a]{raga01a} Raga, A.C., Vel\'azquez, P.F., Cant\'o, J., Masciadri, E. \& Rodr\'\i guez, L.F. 2001a, ApJ, 559, L33.
\bibitem[Raga et al. 2001b]{raga01b} Raga, A.C., Sobral, H., Villagr\'an-Muniz, M., Navarro-Gonz\'alez,
R. and Masciadri, E. 2001b, MNRAS, 324, 206.
\bibitem[Raga, Navarro-Gonz\'alez \& Villagr\'an-Muniz 2000{raga00} Raga, A.C., Navarro-Gonz\'alez, R., Villagr\'an-Muniz, M. 2000,
RMxAA, 36, 67.
\bibitem[Shull \& van Steenberg 1982]{shu82} Shull, J.M. \& van Steenberg, M., 1982, ApJSS, 48, 95.
\bibitem[Sobral et al. 2000]{sobral00} Sobral, H., Villagr\'an-Muniz, M., Navarro-Gonz\'alez, R.
           and Raga A.C. 2000, App. Phys. Lett., 77, 3158.
\bibitem[Van Leer 1982]{vanleer82} Van Leer, B. 1982, ICASE Report N0. 82-30.
\bibitem[Vel\'azquez et al. 2001a]{vel01a} Vel\'azquez, P.F., Sobral, H., Raga, A.C., Villagr\'an-Muniz, M.,
           Navarro-Gonz\'alez, R. 2001a, RMxAA, 37, 87.
\bibitem[Vel\'azquez et al. 2001b]{vel01b} Vel\'azquez, P.F., de la Fuente, E., Rosado, M. and Raga, A.C.
           2001b, A\&A, 377, 1144.
\bibitem[Ye, Turtle \& Kennicutt Jr. 1991]{ye91} Ye, T., Turtle, A.J., \&  Kennicutt Jr., R.C. 1991, MNRAS, 249, 722.
\bibitem[Yusef-Zadeh et al. 2002]{yusef02} Yusef-Zadeh, F, Law, C., Wardle, M, Wang, Q.D., Fruscioni, A., Lang,
C.C. \& Cotera, A., 2002, ApJ, 570, 665.
\end{thebibliography}
\end{document}